\documentclass[11pt]{article}
\usepackage[english]{babel}
\usepackage{cite}
\usepackage{amssymb,amsmath,amsthm,mathrsfs}
\usepackage{xcolor}
\newcommand{\BH}{\mathcal{B}(\mathcal{H})}
\def\<{\langle}
\def\>{\rangle}
\newtheorem{proposition}{Proposition}
\newtheorem{theorem}{Theorem}
\newtheorem{corollary}{Corollary}

\theoremstyle{definition}
\newtheorem{definition}{Definition}

\theoremstyle{remark}
\newtheorem{remark}{Remark}

\def\ri{\mathrm{id}}

\def\bea{\begin{eqnarray}}
\def\eea{\end{eqnarray}}

\newcommand{\bH}{\mathcal{B}(\mathcal{H})}

\begin{document}

\title{\bf Quantum entropy and non-Markovian evolution}

\author{
Paolo Aniello$^{1,2}$, Joonwoo Bae$^3$, and Dariusz Chru\'sci\'nski$^4$
\vspace{2mm}
\\ \small \it $^1$Dipartimento di Fisica ``Ettore Pancini'', Universit\`a di Napoli ``Federico II'',
\\ \small \it Complesso Universitario di Monte S.\ Angelo, via Cintia, I-80126 Napoli, Italy
\vspace{1mm}
\\ \small \it $^2$Istituto Nazionale di Fisica Nucleare, Sezione di Napoli,
\\ \small \it Complesso Universitario di Monte S.\ Angelo, via Cintia, I-80126 Napoli, Italy
\vspace{1mm}
\\ \small \it  $^3$  School of Electrical Engineering, Korea Advanced Institute of Science and Technology (KAIST),  \\ \small \it 291 Daehak-ro Yuseong-gu, Daejeon 34141 Republic of Korea.
\vspace{1mm}
\\ \small \it $^4$Institute of Physics, Faculty of Physics, Astronomy and Informatics,
\\ \small \it Nicolaus Copernicus University,
\\ \small \it Grudziadzka 5, 87100 Toru\'n, Poland
}

\date{}

\maketitle

\begin{abstract}
\noindent
Entropy, and its temporal evolution, play a central role in the foundations of quantum theory and in modern quantum technologies. Here we study, in particular, the relations between the --- in general, non-Markovian --- evolution of an open quantum system, the notions of divisibility of a dynamical map and of distinguishability of quantum states, and the temporal behaviour of various entropy-related quantities such as the R\'enyi (and sandwiched R\'enyi) divergences, and the so-called min- and max- conditional entropies. This, in turn, gives rise to an operational meaning of (non-)Markovianity.
\end{abstract}

%%%------------------------------------------------------------------------------
\section{Introduction}
\label{intro}
%%%------------------------------------------------------------------------------

The description of the evolution of a quantum system interacting with its environment is of fundamental importance.
Indeed, any realistic quantum system interacts, to some extent, with the external world, and suitably describing
the effects of such an interaction is the main aim of the theory of open quantum systems~\cite{open1,open2,open3}.

Over the last decade or so, there has been an increasing interest in open quantum systems, especially in connection with diverse topics
in the broad research area of modern quantum technologies, such as quantum computation, communication, cryptography,
control and metrology~\cite{Rieffel,Wilde,QIT,Sergienko,Wiseman}, to mention just some of the most prominent examples;
in addition, it is also worth mentioning the study of quantum decoherence effects in biology~\cite{Lambert}.

For describing an open quantum system one often adopts the so-called \emph{Markovian} --- or memoryless --- approximation.
Roughly speaking, this amounts to assuming that, when a quantum system interacts with a reservoir (or bath) ---
another, typically much larger, quantum system --- the latter very rapidly looses any memory of the system's past states. From the technical point of view,
this approximation requires a weak coupling between the system and the environment, and
also a suitable separation of the relevant time scales of the system's and environmental evolution~\cite{open1}.

On the other hand, current experimental techniques and technological applications call for a more refined approach
which takes into account certain memory effects --- that are totally neglected in the standard Markovian approximation ---
and hence one has to go beyond the scheme of pure Markovian evolution.
Indeed, recently the issue of non-Markovian evolution of a quantum system has received a considerable attention;
see, e.g., the reviews~\cite{NM1,NM2,NM3}, the recent paper~\cite{WISE} devoted to a comparative analysis of various approaches
to quantum non-Markovian evolution, and the articles~\cite{exp,exp2} reporting on experimental tests.

It should be stressed however that --- whereas, in the classical setting, there is a universal consensus about the notion
of Markovianity for a stochastic process~\cite{Grimmett} --- in the quantum setting there is no universal approach to non-Markovian evolution,
and even the very notion of quantum (non-)Markovianity is not uniquely defined~\cite{NM1,NM2,NM3,WISE}.
Due to the peculiar mathematical description of a quantum system versus a classical one, formalizing in a sensible and rigorous way
this physically intuitive concept seems to be a highly nontrivial task.

Let us recall that the temporal evolution of an open quantum system is usually described by a time-dependent dynamical map
$\Lambda_t \colon \BH \to \BH$, which is completely positive and trace preserving (CPTP)~\cite{open1,open3}.
In this paper, we will consider only finite-dimensional systems, and $\BH$ is just the vector space of
all linear operators acting on the Hilbert space $\mathcal{H}$ associated with a quantum system. Clearly, the dynamical map
satisfies the natural initial condition $\Lambda_0 = {\rm id}$. Any such a map can be obtained via the well known reduction procedure
\begin{equation}
\Lambda_t(\rho) = {\rm Tr}_E ( U_t \rho \otimes \rho_E U_t^\dagger) ,
\end{equation}
where $U_t = e^{-i H t}$, with $H$ being the total system-environment Hamiltonian, and $\rho_E$ is a fixed state of the environment.

With the standard Markovian approximation, one finds out that the dynamical map $\Lambda_t$ satisfies the celebrated
Gorini-Kossakowski-Sudarshan-Lindblad (GKSL) master equation~\cite{GKS,L}, i.e.,
\begin{equation}\label{semi}
\dot{\Lambda}_t = \mathcal{L} \Lambda_t .
\end{equation}
Here the infinitesimal generator $\mathcal{L} : \BH \to \BH$ has the following well known form~\cite{GKS,L}
(also see~\cite{OSID} for the brief history of GKLS master equation):
\begin{equation}\label{}
  \mathcal{L}(\rho) = - i[H_{\rm eff},\rho] + \sum_i \gamma_i \left(V_i \rho V_i^\dagger - \frac 12 \{ V_i^\dagger V_i,\rho\} \right) ,
\end{equation}
where $H_{\rm eff}$ denotes the {\em effective} Hamiltonian of the (reduced) system and the constants $\gamma_i > 0$ are decoherence/dissipation rates.
The corresponding solution for the dynamical map $t\mapsto\Lambda_t = e^{ \mathcal{L} t}$ is a semigroup of operators~\cite{Yosida}, often called a
\emph{Markovian semigroup}~\cite{Alicki}. This case provides the simplest and mathematically best-behaved --- but, of course, most restrictive ---
notion of quantum Markovianity.

To go beyond the semigroup master equation, one can use either a time-dependent generator $\mathcal{L}_t$,
or the so-called \emph{memory kernel master equation}
\begin{equation}\label{semi}
  \dot{\Lambda}_t = \int_0^t \mathcal{K}_{t-\tau} \Lambda_\tau d\tau ,
\end{equation}
where the kernel $\mathcal{K}_t : \BH \to \BH$ takes into account non-negligible memory effects induced by the interaction of the system with the environment.
On the other hand, undertaking this more sophisticated approach, it is highly nontrivial to find necessary and sufficient conditions for the integral kernel
to give rise to a CPTP dynamical map $\Lambda_t$ (see, e.g., \cite{kernel1,kernel2}).

In this context, interesting and less restrictive characterizations of non-Markovianity of an open quantum system evolution
are based on two fundamental properties: the distinguishability of physical states~\cite{BLP} and the CP-divisibility of a quantum dynamical map~\cite{RHP}
(see also~\cite{Wolf1,Wolf2,Wissmann}).
But many other proposals can be found in the literature, including approaches relying on quantum Fisher information~\cite{F}, quantum fidelity~\cite{Fi},
mutual information~\cite{Luo}, channel capacity~\cite{Bogna}, geometry of the set of accessible
states~\cite{Geo},  quantum interferometric power~\cite{Power} (see the recent reviews~\cite{NM1,NM2,NM3,WISE}, for a more exhaustive list).

In the present paper --- rather than trying to trace a new border between the Markovian and non-Markovian quantum regimes ---
we mainly aim at exploring the relations connecting the notion of quantum (non-)Markovianity with various entropic quantities.
This may provide new insights, because entropy is one of the most fundamental concepts in physics and information theory~\cite{Wehrl,O-P, Sha1,Sha2},
and a profound and meaningful link between the two disciplines. We will also try to set connections between various scattered
results in the literature.

The paper is organized as follows. We start, in the next section, by recalling the two aforementioned influential approaches to quantum Markovianity,
based on divisibility of quantum dynamical maps and distinguishability of quantum states (density operators), respectively.
We stress that the divisibility property of dynamical maps is related to the data processing inequalities
which play an essential role in quantum information theory~\cite{Wilde,QIT,Holevo-2012}. On the other hand, divisibility is connected to
distinguishability of states (Section~\ref{STATES}) and to discrimination of quantum channels (Section~\ref{CHANNELS}).
In Section~\ref{ENTROPY}, we relate Markovianity to relative entropy, and to relative R\'enyi entropy of order $\alpha$.
These results are then generalized to an important family of sandwiched R\'enyi divergences; see Section~\ref{RENYI}.
It turns out that the sandwiched R\'enyi divergences give rise to min- and max- conditional entropies with a clear
operational meaning (Section~\ref{OP}). Conclusions and final remarks are collected in Section~\ref{CON}.

\section{Markovianity and divisibility} \label{STATES}

A dynamical map $\Lambda_t : \BH \to \BH$ is a family of completely positive trace-preserving (CPTP) maps acting on space $\BH$ supplemented by the natural initial condition $\Lambda_0 = {\rm id}$. Recall, that a linear map $\Phi : \BH \to \BH$ is $k$-positive if ${\rm id}_k \otimes \Phi : \mathbb{M}_k(\BH) \to \mathbb{M}_k(\BH)$ is positive ($\mathbb{M}_k(\BH)$ is an algebra of $k\times k$ matrices with entries from $\BH$). A map $\Phi$ which is $k$-positive for $k=1,2,\ldots$ is called completely positive. It was shown by Choi \cite{Choi} that $\Phi$ is completely positive if and only if it is $d$-positive, where $d = {\rm dim}\, \mathcal{H}$ (see \cite{Paulsen} for more mathematical details).  One calls $\Lambda_t$ \emph{divisible} if it can be decomposed as
\begin{equation}\label{div}
  \Lambda_t = V_{t,s} \Lambda_s,
\end{equation}
where  $V_{t,s}:\BH \rightarrow \BH$  for any $t \geq s$. Note, that if $\Lambda_t$ is invertible  for all $t\geq 0$, then it is trivially divisible since $V_{t,s} = \Lambda_t \Lambda_s^{-1}$.   One calls $\Lambda_t$ $k$-divisible \cite{PRL-k} if $V_{t,s}$ is $k$-positive.  In particular  $1$-divisible maps are called P-divisible and $d$-divisible maps CP-divisible.  Following \cite{RHP} one can find the definition of Markovianity as follows.

\begin{definition}[\cite{RHP}]
A dynamical map $\Lambda_t$ represents a Markovian evolution if $\Lambda_t$ is CP-divisible.
\end{definition}

In  Ref. \cite{BLP}, Breuer et al proposed another approach to define Markovianity, based on the distinguishability of quantum states under a quantum dynamics. Namely, the information flow of a dynamical map $\Lambda_t$ has been introduced for a pair of initial states $\rho_1$ and $\rho_2$,
\begin{equation}\label{inf-flow}
\sigma(\rho_1,\rho_2;t) =  \frac{d}{dt} ||\Lambda_t(\rho_1-\rho_2)||_1 ,
\end{equation}
where $||A||_1$ for $A \in \BH$ denotes the trace-norm, that is, $||A||_1 = {\rm Tr}|A|$. If the distinguishability increases in time, it may imply that the system gains information from environment, called information backflow, that indicates non-Markovianity. The definition of quantum Markov evolution has been then introduced as follows.

\begin{definition}[\cite{BLP}] A dynamical map $\Lambda_t$ represents a Markovian evolution if
\begin{equation}\label{flow}
  \sigma(\rho_1,\rho_2;t) \leq 0 ,
\end{equation}
for any pair of quantum states $\{ \rho_1,\rho_2\}$ and $t\geq 0$.
\end{definition}

These two very influential concepts are not independent and the intricate relation between them was analyzed \cite{Angel}. Let us consider a general problem of distinguishability of the states from the ensemble $\{p_i,\rho_i\}_{i=1}^N$, that describes preparation of state $\rho_{i}$ with probability $p_i$. When one of the states is given, the problem of optimal state discrimination aims to find optimal measurement, that is, a set of positive-operator-valued-measures (POVMs), that maximizes the average probability of making a correct guess, called the guessing probability \cite{H-73}. Let $\{ E_j\}_{j=1}^N$ denote POVMs. The guessing probability is given as follows,
 \begin{equation}\label{GUESS}
  p_{\rm guess}(\{p_i,\rho_i\}_{i=1}^N) = \max_{ \{ {E}_i\}_{j=1}^N} \sum_i p_i {\rm Tr}(E_i \rho_i) ,
\end{equation}
where we maximize over all POVMs $\{E_i\}$. This definition immediately implies the following, namely that the guessing probability does not increase under a quantum channel.

\begin{proposition} If $\Phi : \bH \to \BH$ is a positive trace-preserving map, then
\begin{equation}\label{}
   p_{\rm guess}(\{p_i,\rho_i\}_{i=1}^N) \geq  p_{\rm guess}(\{p_i,\Phi(\rho_i)\}) ,
\end{equation}
for any ensemble $\{p_i,\rho_i\}_{i=1}^N$.
\end{proposition}

This can be applied to the case that the channel is P-divisible, as follows.

\begin{proposition} If the dynamical map $\Lambda_t$ is P-divisible, then
\begin{equation}\label{}
   \frac{d}{dt} p_{\rm guess}(\{p_i,\Lambda_t(\rho_i)\}) \leq  0 ,
\end{equation}
for any ensemble $\{p_i,\rho_i\}_{i=1}^N$.
\end{proposition}

Similarly, if the system is coupled to $k$-dimensional ancilla and one considers the ensemble $\{p_i,\widetilde{\rho}_i\}_{i=1}^N$, where $\widetilde{\rho}_i$ are states living in $\mathcal{H} \otimes \mathbb{C}^k$, then

\begin{proposition} \label{P-3} If the dynamical map $\Lambda_t$ is  $k$-divisible, then
\begin{equation}\label{}
   \frac{d}{dt} p_{\rm guess}(\{p_i,\Lambda_t(\widetilde{\rho}_i)\}_{i=1}^N) \leq  0 ,
\end{equation}
for any ensemble $\{p_i,\widetilde{\rho}_i\}_{i=1}^N$.
\end{proposition}

Although the general form the guessing probability is formulated, its closed form is known so far for limited cases, see the recent results \cite{ref:bae1, ref:bae2}. For $N=2$, the guessing probability in Eq. (\ref{GUESS}) has the closed form as follows,
\begin{equation}\label{}
   p_{\rm guess} = \frac 12 \left(1+ ||p_1 \rho_1 - p_2 \rho_2||_1 \right) , \nonumber
\end{equation}
It is found that the trace norm of the operator $X = p_1 \rho_1 - p_2 \rho_2$ determines the guessing probability, where $X$ is called the Helstrom matrix \cite{Helstrom}. Hence it is clear that the condition (\ref{flow}) is an immediate consequence of P-divisibility. Interestingly, the result has been extended to discrimination with $k$-dimensional ancillary systems as follows.

\begin{theorem}[\cite{Angel,PRL-k}] Suppose that the map $\Lambda_t$ is invertible. Then it is $k$-divisible if and only if
  \begin{equation}\label{k-div}
    \frac{d}{dt} || [{\rm id}_k \otimes \Lambda_t](p_1 \widetilde{\rho}_1 - p_2 \widetilde{\rho}_2)||_1  \leq 0 ,
  \end{equation}
for any pair of quantum states $\{p_i,\widetilde{\rho}_i \}_{i=1}^2$.
\end{theorem}

To find Markovianity, in fact it suffices to consider a pair of quantum states that appear with equal probability, i.e., $p_1=p_2=\frac 12$, while the ancillary Hilbert space enlarged, as follows.

\begin{theorem}[\cite{BOGNA}] Suppose that the map $\Lambda_t$ is invertible. Then it is CP-divisible if and only if
  \begin{equation}\label{k-div}
    \frac{d}{dt} || [{\rm id}_{d+1} \otimes \Lambda_t]( \widetilde{\rho}_1 -  \widetilde{\rho}_2)||_1  \leq 0 ,
  \end{equation}
for any ensemble pair of initial states $\{\widetilde{\rho}_1, \widetilde{\rho}_2\}$ living in $\mathcal{H} \otimes \mathbb{C}^{d+1}$, where the dimension of system Hilbert space ${\rm dim}\mathcal{H} =d$.
\end{theorem}

The problem of maps which are not invertible was recently considered in \cite{BOGNA} and \cite{Erling}. Buscemi and Datta \cite{Datta} analyzed a similar problem but for the discrete time evolution $\Lambda_k$ $(k=0,1,2,\ldots)$. In this scenario CP-divisibility is realized by CPTP maps $V_{i,j}$ for any pair $i > j$. They proved that CP-divisibility is equivalent to
\begin{equation}\label{BD}
  p_{\rm guess}(\{p_i,[{\rm id}\otimes \Lambda_k](\widetilde{\rho}_i)\}) \leq  p_{\rm guess}(\{p_i,[{\rm id}\otimes \Lambda_l](\widetilde{\rho}_i)\}) ,
\end{equation}
for any pair $k > l$, any ensemble $\{p_i,\widetilde{\rho}_i\}_{i=1}^n$, where $\widetilde{\rho}_i$ are states living in $\mathcal{H} \otimes \mathcal{H}$. This result applies for arbitrary dynamical maps (not necessarily invertible) but requires ensembles with arbitrary large number of elements.

\section{Entropic criteria of non-Markovianity} \label{ENTROPY}

In this section, we recall some basic facts about quantum entropies to be used for the analysis of Markovianity
in the following sections. One of the basic quantities is a relative entropy which is then used  to define other entropic quantities. For any pair of quantum states $\rho$ and $\sigma$ one defines the quantum relative entropy (see e.g. \cite{Wehrl,O-P})

\begin{equation}\label{Relative}
  D(\rho||\sigma) = \left\{ \begin{array}{ll} {\rm Tr}[\rho(\log\rho - \log\sigma])\ , & \ \mbox{if}\ {\rm supp}\, \rho \subseteq {\rm supp}\, \sigma \\ + \infty \ , & \ \mbox{otherwise} \end{array} \right. \ .
\end{equation}
This quantity was introduced by Umegaki as a generalization of the classical Kullback-Leibler divergence. Lindblad \cite{Lindblad} and Uhlmann \cite{Uhlmann} proved that $D(\rho||\sigma) $ satisfies the data processing inequality, that is, for any CPTP map $\mathcal{E} : \BH \to \BH$
\begin{equation}\label{DPI}
   D(\rho||\sigma) \geq  D(\mathcal{E}(\rho)||\mathcal{E}(\sigma)) .
\end{equation}
Actually, Uhlmann \cite{Uhlmann} {has shown} that it is sufficient that the map $\mathcal{E}$ is trace-preserving and 2-positive
(a 2-positive maps is also called coarse graining \cite{Petz}). Recently this result was generalized in \cite{Reeb} to arbitrary positive trace-preserving maps. It implies the following

\begin{proposition} If the dynamical map $\Lambda_t$ is $k$-divisible, then
  \begin{equation}\label{}
    \frac{d}{dt} D(\Lambda_t(\widetilde{\rho})||\Lambda_t(\widetilde{\sigma}))\leq 0,
  \end{equation}
where $\widetilde{\rho}$ and $\widetilde{\sigma}$ are density operators living in $\mathbb{M}_k(\BH)$.
\end{proposition}

There has been the definition of the whole family of quantum R\'enyi-$\alpha$ divergences via \cite{O-P}
\begin{equation}\label{}
  D_\alpha(\rho||\sigma) = \frac{1}{\alpha -1} \log {\rm Tr}[\rho^\alpha \sigma^{1-\alpha}] ,
\end{equation}
for $\alpha \in (0,1) \cup (1,+\infty)$. One recovers the standard relative entropy (\ref{Relative}) in the limit for $\alpha \to 1$:
$\lim_{\alpha \to 1}  D_\alpha(\rho||\sigma) =  D(\rho||\sigma)$. R\'enyi-$\alpha$ divergence is monotonic under CPTP maps (i.e.,
satisfies the data processing inequality) for $\alpha \in (0,1) \cup (1,2]$ \cite{O-P,Hiai-Petz} (and in the $\alpha \to 1$ limit).
{ Moreover, it is known that it is monotonic under positive trace-preserving maps in the limit for $\alpha \to 0$ and for $\alpha \to 1$ (cf. \cite{Reeb});
the problem is open for $\alpha \in (0,1) \cup (1,2)$ \cite{Hiai-Petz}. Setting $D_{\alpha_0}(\rho||\sigma)=\lim_{\alpha \to \alpha_0}  D_\alpha(\rho||\sigma)$,
for $\alpha_0\in \{0,1\}$, these results imply the following.
}

\begin{proposition}  If the dynamical map $\Lambda_t$ is CP-divisible (i.e. Markovian), then
  \begin{equation}\label{}
    \frac{d}{dt} D_\alpha(\Lambda_t({\rho})||\Lambda_t({\sigma}))\leq 0,
  \end{equation}
for $\alpha \in (0,1) \cup (1,2]$.    If the dynamical map $\Lambda_t$ is $k$-divisible, then
 \begin{equation}\label{}
    \frac{d}{dt} D_\alpha([{\rm id}_k \otimes\Lambda_t](\widetilde{\rho})\,||\,{\rm id}_k \otimes\Lambda_t](\widetilde{\sigma}))\leq 0,
  \end{equation}
for $\rho,\sigma \in \mathbb{M}_k(\BH)$ and $\alpha \in \{0,1,2\}$.
\end{proposition}
{ Note that if $\sigma$ is the maximally mixed state --- i.e., $\sigma=\rho_\star\equiv\mathbb{I}/d$ --- then
\begin{equation}\label{}
    D_\alpha(\rho||\sigma) = - S_\alpha(\rho) + \log d ,
\end{equation}
where $S_\alpha(\rho) = \frac{1}{1-\alpha} \log {\rm Tr}\rho^\alpha$ stands for the R\'enyi entropy of order $\alpha$, for
$\alpha\in(0,1)\cup(1,+\infty)$, and $S_{1}(\rho)=\lim_{\alpha \to 1}  S_\alpha(\rho)$.
Therefore, using results from the recent papers~\cite{Paolo,Paolo-bis},
one can easily derive the following }

\begin{proposition} {
Let $t\mapsto\Lambda_t$ be P-divisible and $\alpha\in(0,+\infty)$. If $\Lambda_t$ is unital $(t\ge 0)$, then,
for every state $\rho$,
\begin{equation}
D_\alpha(\Lambda_t(\rho)||\Lambda_t(\rho_\star)) = D_\alpha(\Lambda_t(\rho)||\rho_\star)
= -S_\alpha(\Lambda_t(\rho)) + \log d
\end{equation}
and
\begin{equation}
\frac{d}{dt} D_\alpha(\Lambda_t(\rho)||\rho_\star) \le 0 .
\end{equation}
On the other hand, if, for every state $\rho$,
\begin{equation}
\frac{d}{dt} D_\alpha(\Lambda_t(\rho)||\rho_\star) \le 0 ,
\end{equation}
then $\Lambda_t$ is unital $(t\ge 0)$, so that
$D_\alpha(\Lambda_t(\rho)||\Lambda_t(\rho_\star)) = D_\alpha(\Lambda_t(\rho)||\rho_\star) = -S_\alpha(\Lambda_t(\rho)) + \log d$.
}
\end{proposition}

{
\begin{proof}
Notice that $S_\alpha(\rho)$ is of the form $f_\alpha(\rho)$, where $f_\alpha$ is a Schur-concave function on the convex set
of states, which attains a strict global maximum at $\rho=\rho_\star$; see sect.~{2} of~\cite{Paolo}
(also see remarks~{9} and~{10} \emph{ibidem}), or propositions~{2} and~{3} of~\cite{Paolo-bis}.
Therefore, by lemma~{1} of~\cite{Paolo}, and taking into account remark~{8} \emph{ibidem}, $S_\alpha(\rho)$ is not decreased
by a positive trace-preserving map if and only if this is unital. The statement follows.
\end{proof}
}

\section{Sandwiched R\'enyi divergences and min- and max-entropies} \label{RENYI}

Recently, a new class of R\'enyi-$\alpha$ divergences was introduced \cite{Tom,WWY}. They are called {\em sandwiched R\'enyi divergences} and defined via
\begin{equation}\label{Sand}
   \widetilde{D}_\alpha(\rho||\sigma) =  \frac{1}{\alpha-1} \log \left( {\rm Tr}\left[ \left( \sigma^{\frac{1-\alpha}{2\alpha}} \rho \, \sigma^{\frac{1-\alpha}{2\alpha}} \right)^\alpha \right] \right) ,
\end{equation}
if $\,{\rm supp}\, \rho \subseteq {\rm supp}\, \sigma$, and equals $+\infty$ otherwise. Clearly, if $\rho$ and $\sigma$ commute, then $ \widetilde{D}_\alpha(\rho||\sigma) =  {D}_\alpha(\rho||\sigma)$. Interestingly one has the following relation between $D_\alpha$ and $\widetilde{D}_\alpha$ \cite{TOM}
\begin{equation}\label{}
  \widetilde{D}_\alpha(\rho||\sigma) = \lim_{n\to \infty}\, \frac 1n {D}_\alpha(\mathcal{P}_{\sigma^{\otimes n}}(\rho^{\otimes n})\,||\,\sigma^{\otimes n}) ,
\end{equation}
where the quantum channel $\mathcal{P}_\sigma$ is defined by $\mathcal{P}_\sigma(X) = \sum_k P_k X P_k$, and $\sigma = \sum_k \lambda_k P_k$ stands for the spectral decomposition of $\sigma$, that is, $\mathcal{P}_\sigma$ decoheres w.r.t. eigenbasis of $\sigma$.

Again, one recovers standard von Neumann relative entropy as the $\alpha \to 1$ limit of the sandwiched R\'enyi divergences.
Moreover
\begin{equation}\label{}
  \widetilde{D}_{1/2}(\rho||\sigma) = -2 \log F(\rho,\sigma) ,
\end{equation}
where $F(\rho,\sigma) = || \sqrt{\rho} \sqrt{\sigma}||_1$ is the quantum fidelity. The sandwiched R\'enyi divergences enjoy many interesting properties (cf. the recent monograph \cite{TOM}). For our purposes the central property is the data processing inequality.

\begin{theorem}[\cite{Lieb,Tom,Beigi,Ogawa}] \label{Ogawa} For an arbitrary quantum channel $\mathcal{E}$ the sandwiched R\'enyi relative entropy of order $\alpha$ satisfies the data processing inequality
\begin{equation}\label{Data-CP}
  \widetilde{D}_\alpha(\mathcal{E}(\rho)||\mathcal{E}(\sigma))    \leq \widetilde{D}_\alpha(\rho||\sigma)  ,
\end{equation}
for $\alpha \in [\frac 12,1) \cup (1,\infty)$.
\end{theorem}

The theorem can be applied to CP-divisible maps as follows.

\begin{corollary} If $\Lambda_t$ is CP-divisible, then
\begin{equation}\label{}
  \frac{d}{dt}  \widetilde{D}_\alpha(\Lambda_t(\rho) || \Lambda_t(\sigma)) \leq 0 ,
\end{equation}
for $\alpha \in [\frac 12,1) \cup (1,\infty)$. In particular for $\alpha = 1/2$ one has
\begin{equation}\label{}
  \frac{d}{dt}  F(\Lambda_t(\rho),\Lambda_t(\sigma)) \geq 0 ,
\end{equation}
for any CP-divisible map.
\end{corollary}
Note, that quantum fidelity $F(\rho,\sigma)$ is equivalent to the trace distance $||\rho-\sigma||_1$ by means of the following relation \cite{QIT}

\begin{equation}\label{<<}
  1 - F(\rho,\sigma) \leq \frac 12 ||\rho-\sigma||_1 \leq \sqrt{  1 - F^2(\rho,\sigma) } ,
\end{equation}
for any pair of states $\rho$ and $\sigma$.  Interestingly, in a recent paper \cite{Reeb} Theorem \ref{Ogawa}  was extended to positive trace-preserving maps

\begin{theorem}[\cite{Reeb}] For an arbitrary positive trace-preserving map $\Phi$ the sandwiched R\'enyi relative entropy of order $\alpha$ satisfied data processing inequality
\begin{equation}\label{Data-P}
  \widetilde{D}_\alpha(\Phi(\rho)||\Phi(\sigma))    \leq \widetilde{D}_\alpha(\rho||\sigma)  ,
\end{equation}
for $\alpha \in \{\frac 12\} \cup (1,\infty)$.
\end{theorem}
The case $\alpha = \frac 12$ was already proved in \cite{Ogawa} and the problem is open for $\alpha \in (\frac 12,1)$ (cf. \cite{Reeb}). The sandwiched R\'enyi divergence gives rise to the very useful concept of conditional entropy.

\begin{corollary} If $\Lambda_t$ is P-divisible, then
\begin{equation}\label{}
  \frac{d}{dt}  \widetilde{D}_\alpha(\Lambda_t(\rho) || \Lambda_t(\sigma)) \leq 0 ,
\end{equation}
for $\alpha \in \{\frac 12\} \cup (1,\infty)$. In particular for $\alpha = 1/2$ one has
\begin{equation}\label{}
  \frac{d}{dt}  F(\Lambda_t(\rho),\Lambda_t(\sigma)) \geq 0 ,
\end{equation}
for any P-divisible map.
\end{corollary}

The standard von Neumann conditional entropy for a bipartite state $\rho_{AB}$ living in $\mathcal{H}_A \otimes \mathcal{H}_B$  is defined by \cite{O-P,Wehrl}
\begin{equation}\label{}
  H(A|B)_\rho = S(\rho_{AB})_\rho - S(\rho_B) ,
\end{equation}
where $\rho_B = {\rm Tr}_A \rho_{AB}$ is a marginal state. This definition is not very convenient in order to generalize conditional entropy for R\'enyi entropy. However, one may equivalently define $H(A|B)_\rho$ as follows
\begin{equation}\label{d}
  H(A|B)_\rho = - \inf_{\sigma_B} D(\rho_{AB} || \mathbb{I}_A \otimes \sigma_B) ,
\end{equation}
where the infimum is taken over all local states $\sigma_B$.

\begin{remark} Note, that in (\ref{d}) we use $\mathbb{I}_A$ which is not a state (not normalized). However, both $D(\rho||\sigma)$ and $\widetilde{D}_\alpha(\rho||\sigma)$ may be generalized if we replace $\rho$ and $\sigma$ by positive operators (not normalized) (cf. \cite{TOM}).
\end{remark}
Now, we may generalize (\ref{d}) to introduce the whole family of conditional entropies: for a bipartite state living in $\mathcal{H}_A \otimes \mathcal{H}_B$ the conditional R\'enyi entropy is defined as follows \cite{Renner-09}

\begin{equation}\label{}
  \widetilde{H}_\alpha(A|B)_\rho = - \inf_{\sigma_B}  \widetilde{D}_\alpha(\rho_{AB} || \mathbb{I}_A \otimes\sigma_B) ,
\end{equation}
where the infimum is over all states $\sigma_B$ of the $B$ subsystem.

\begin{proposition}[\cite{TOM}] Conditional R\'enyi entropy satisfies the following data processing inequality: let $\mathcal{E} : \mathcal{B}(\mathcal{H}_B) \to \mathcal{B}(\mathcal{H}_{B'})$ be a quantum channel. Then for any $\alpha \geq \frac 12$
\begin{equation}\label{}
  \widetilde{H}_\alpha(A|B)_\rho \leq \widetilde{H}_\alpha(A|B')_{\rho'} ,
\end{equation}
where $\rho'_{AB'} = [{\rm id}_A \otimes \mathcal{E}](\rho_{AB})$.
\end{proposition}
Note also that min- and max entropies can be introduced as follows: ${H}_{\rm min}(A|B)_\rho = \lim_{\alpha \to \infty} \widetilde{H}_\alpha(A|B)_\rho$ and ${H}_{\rm max}(A|B)_\rho = \lim_{\alpha \to 1/2} \widetilde{H}_\alpha(A|B)_\rho$. The single-shot entropies can be defined equivalently as follows.

%\begin{equation}\label{MIN}
%{H}_{\rm min}(A|B)_\rho = \lim_{\alpha \to \infty} \widetilde{H}_\alpha(A|B)_\rho ,
%\end{equation}
%and conditional max-entropy
%\begin{equation}\label{MAX}
%{H}_{\rm max}(A|B)_\rho = \lim_{\alpha \to \frac 12} \widetilde{H}_\alpha(A|B)_\rho .
%\end{equation}

\begin{proposition}[\cite{TOM}] Min- and max-entropies satisfy
\begin{equation}\label{MIN-1}
{H}_{\rm min}(A|B)_\rho = \min_{\sigma_B} - \log\, || \mathbb{I}_A \otimes B^{-\frac 12} \, \rho_{AB} \, \mathbb{I}_A \otimes B^{-\frac 12} ||_\infty ,
\end{equation}
and
\begin{equation}\label{MAX-1}
{H}_{\rm max}(A|B)_\rho = \max_{\sigma_B} \log F(\rho_{AB},\mathbb{I}_A \otimes \sigma_B) ,
\end{equation}
where now $\sigma_B \geq 0$ and ${\rm Tr}\sigma_B \leq 1$ (subnormalized).
\end{proposition}

It is worth to note that min- and max-entropies are related by the following relation \cite{Renner-09}
\begin{equation}\label{eq:max}
  H_{\rm min}(A|B)_\rho +  H_{\rm max}(A|C)_\rho = 0 ,
\end{equation}
for arbitrary pure three-partite state $\rho_{ABC}$.  {Min- and max- entropies are useful theoretical tools to provide information-theoretic meanings to operational tasks in a single-shot scenario, such as data compression, channel coding, privacy amplification, state merging, decoupling, etc. \cite{TOM}. }

\section{Operational meaning of min- and max- conditional entropies}   \label{OP}
\label{sec:minmax}

Min- and max- entropies have interesting operational meaning \cite{Renner-09}. Consider the maximally entangled state in $\mathcal{H}_A \otimes \mathcal{H}_B$
\begin{equation}\label{}
  |\psi^+_{AB}\> = \frac{1}{\sqrt{d_A} } \sum_{i=1}^{d_A} |e^A_i\> \otimes |e^B_i\> ,
\end{equation}
where we assumed that ${\rm dim}\mathcal{H}_A \leq {\rm dim}\mathcal{H}_B$. One defines \cite{Renner-09} quantum correlation
\begin{eqnarray}\label{q-corr}
  q_{\rm corr}(A|B)_\rho &=& d_A \, \max_{\mathcal{E}} \, F([{\rm id}_A \otimes \mathcal{E}](\rho_{AB}), |\psi^+_{AB}\>\< \psi^+_{AB}|)^2  \nonumber \\ &=&
  d_A  \, \max_{\mathcal{E}} \, \< \psi^+_{AB}|\, [{\rm id}_A \otimes \mathcal{E}](\rho_{AB})\, |\psi^+_{AB}\>^2 ,
\end{eqnarray}
where the maximum is performed w.r.t. to all quantum channels acting on the $B$ subsystem.  This quantity measures the maximal overlap
with the maximally entangled state $|\psi^+_{AB}\>$ which can be achieved by local quantum channels $\mathcal{E}$. Similarly, one defines the decoupling accuracy
\begin{equation}\label{q-decpl}
  q_{\rm decpl}(A|B)_\rho = d_A \, \max_{\sigma_B} \, F(\rho_{AB}, \mathbb{I_A}/d_A \otimes \sigma_B)^2 ,
\end{equation}
where the maximum is performed over all states $\sigma_B$ of the $B$ subsystem.

\begin{proposition}[\cite{Renner-09}]  The quantum correlation and decoupling accuracy satisfy
\begin{equation}\label{}
   q_{\rm corr}(A|B)_\rho = 2^{-{H}_{\rm min}(A|B)_\rho}\ , \ \ \   q_{\rm decpl}(A|B)_\rho = 2^{{H}_{\rm max}(A|B)_\rho} ,
\end{equation}
respectively.
\end{proposition}

Since max- and min-entropies are dual each other under purification, in what follows we focus on the conditional min-entropy by which the max-entropy straightforwardly follows from the purification. Since $\widetilde{H}(A|B)_\rho$ satisfies data processing inequality one has the following

\begin{proposition} If the dynamical map is CP-divisible, then
\begin{equation}\label{}
  \frac{d}{dt}  q_{\rm corr}(A|B)_{\rho_t} \leq 0 ,
\end{equation}
where $\rho_t =   [{\rm id} \otimes \Lambda_t](\rho_{AB})$, and $\rho_{AB}$ is an arbitrary state in $\mathcal{H} \otimes \mathcal{H}$.
\end{proposition}
This result may be generalized as follows: a density operator $\rho$ living in $\mathcal{H} \otimes \mathcal{H}$ has Schmidt number not greater than $k$ if and only if  $[{\rm id} \otimes \Phi](\rho)  \geq 0$ for all $k$-positive maps $\Phi : \BH \to \BH$. Hence

\begin{proposition} \label{P-11}If the dynamical map is $k$-divisible, then
\begin{equation}\label{}
  \frac{d}{dt}  q_{\rm corr}(A|B)_{\rho_t} \leq 0 ,
\end{equation}
where $\rho_t =   [{\rm id} \otimes \Lambda_t](\rho_{AB})$, and $\rho_{AB}$ is an arbitrary state in $\mathcal{H} \otimes \mathcal{H}$ with Schmidt number not greater than $k$.
\end{proposition}
Note, that the above results refine the original idea of \cite{RHP}: if $\mathcal{M}$ is a genuine entanglement measure, then if $\Lambda_t$ is CP-divisible one has
\begin{equation}\label{EEE}
  \frac{d}{dt} \mathcal{M}([{\rm id} \otimes \Lambda_t](\rho_{AB}) ) \leq 0 ,
\end{equation}
for all bipartite states living in $\mathcal{H} \otimes \mathcal{H}$. If $\Lambda_t$ is only $k$-divisible, then (\ref{EEE}) holds for all states with Schmidt number not greater than $k$.

Now, let us suppose that $\rho_{AB}$ is a classical-quantum state
\begin{equation}\label{}
  \rho_{AB} = \sum_i p_i |e^A_i\>\<e^A_i| \otimes \rho^B_i.
\end{equation}
It turns out that the quantum correlation in Eq. (\ref{q-corr}) corresponds to the guessing probability,
\begin{equation}\label{}
  q_{\rm corr}(A|B)_\rho = p_{\rm guess}(\{p_i,\rho^B_i\}) ,
\end{equation}
where the guessing probability is defined in (\ref{GUESS}) and hence in this case Proposition \ref{P-11}   implies Proposition \ref{P-3}.

\section{Markovianity and channel discrimination}  \label{CHANNELS}
%with one-shot entropic quantities}

Given an ensemble od quantum channels $\mathcal{E}_k: \BH \to \BH$ with probabilities $p_k$ one faces the discrimination problem similar to that of discriminating quantum states from an ensemble $\{ p_k,\rho_k\}$. The effectiveness of discrimination depends upon the resources at our disposal. Coupling the system to the $k$-dimensional ancilla one defines the guessing probability
\begin{equation}\label{}
  p_{\rm guess}^{(k)}(\{p_i,\mathcal{E}_i\}) = \max_{\widetilde{\rho_k}} p_{\rm guess}(\{p_i,[{\rm id}_k \otimes \mathcal{E}_i](\widetilde{\rho}_k) \}) ,
\end{equation}
where $\widetilde{\rho}_k \in \mathbb{M}_k(\BH)$. Equivalently,
\begin{equation}\label{}
  p_{\rm guess}^{(k)}(\{p_i,\mathcal{E}_i\}) = \max_{\widetilde{\rho}} p_{\rm guess}(\{p_i,[{\rm id} \otimes \mathcal{E}_i](\widetilde{\rho}) \}) ,
\end{equation}
where $\widetilde{\rho}$ is a state in $\mathcal{H} \otimes \mathcal{H}$ with Schmidt number not greater than $k$. It is clear that

\begin{equation}\label{}
   p_{\rm guess}^{(1)}(\{p_i,\mathcal{E}_i\}) \leq  p_{\rm guess}^{(2)}(\{p_i,\mathcal{E}_i\}) \leq \ldots \leq  p_{\rm guess}^{(d)}(\{p_i,\mathcal{E}_i\}) ,
\end{equation}
where $d = {\rm dim}\mathcal{H}$. Again, for two channels $\mathcal{E}_1$ and $\mathcal{E}_2$ with probabilities $p_1=p$, and $p_2=1-p$, respectively, one finds
\begin{equation}\label{}
p_{\rm guess}^{(k)}(\{p_i,\mathcal{E}_i\}) = \frac 12 \left( 1 + D_1^p(\mathcal{E}_1,\mathcal{E}_2) \right) ,
\end{equation}
where we introduced the distance $D_k^p$ in the space of linear maps $\mathcal{L}(\BH,\BH)$:
\bea
D_{k}^{p} (\mathcal{E}_{1} , \mathcal{E}_{2} ) =  \| \ri_k \otimes (  (1-p) \mathcal{E}_1 -p ~\mathcal{E}_2)  \|_1 ,
\eea
and $||\Phi||_1 = \max_{\widetilde{\rho}} ||\Phi(\widetilde{\rho})||_1$, where $\widetilde{\rho}$ are density operators living in $\mathcal{H} \otimes \mathbb{C}^k$. One has
\begin{equation}\label{}
  D_1^p(\mathcal{E}_1,\mathcal{E}_2) \leq D_2^p(\mathcal{E}_1,\mathcal{E}_2) \leq \ldots \leq D_d^p(\mathcal{E}_1,\mathcal{E}_2) .
\end{equation}
Note, that the last `$d$-distance is defined in terms of the so called diamond norm \cite{Watrous}
\begin{equation}\label{}
  || \Phi||_\diamond := ||{\rm id} \otimes \Phi||_1 = \sup_{||\rho||_1 \leq 1} ||[{\rm id} \otimes \Phi](\rho)||_1
\end{equation}
In the Heisenberg picture $\Phi^\#$ it is related to the norm of complete boundness (cb-norm) \cite{Paulsen}
\begin{equation}\label{}
  || \Phi^\#||_{\rm cb} = ||{\rm id} \otimes \Phi^\#||_\infty =\sup_{||X||_\infty \leq 1} ||[{\rm id} \otimes \Phi^\#](X)||_\infty
\end{equation}
where $|| X ||_\infty$ stands for the operator norm, that is, $||\Phi||_\diamond = ||\Phi^\#||_{\rm cb}$.   In a recent paper \cite{Gross} the authors introduced so called square norm of a bipartite operator $X \in \mathcal{B}(\mathcal{H}_A \otimes \mathcal{H}_B)$:
\begin{equation}\label{}
  || X ||_\Box = \sup_{B_1,B_2 \in \mathcal{B}(\mathcal{H}_B)} \left\{ \, || (\mathbb{I}_A \otimes B_1) \, X \, (\mathbb{I}_A \otimes B_2)||_1 \  ; \ ||B_1||_2=||B_2||_2 = \sqrt{d_B} \, \right\} ,
\end{equation}
with $d_B = {\rm dim} \mathcal{H}_B$. One shows \cite{Gross} that for a linear map $\Phi : \mathcal{B}(\mathcal{H}_B) \to \mathcal{B}(\mathcal{H}_A)$ one has

\begin{equation}\label{}
  ||  J(\Phi) ||_\Box = d_B || \Phi ||_\diamond ,
\end{equation}
where $J(\Phi) = \sum_{i,j=1}^{d_B} \Phi(|i\>_B\<j|) \otimes |i\>_B\<j|$ is the corresponding Choi matrix.

\begin{proposition}[\cite{Angel}] For any pair of channels $\mathcal{E}_1$ and $\mathcal{E}_2$, let us define $\Delta =  p_1 \mathcal{E}_1 - p_2 \mathcal{E}_2$ and
\begin{equation}\label{}
  X = \sum_{i,j=1}^d |i\>\<j| \otimes \Delta(|i\>\<j|). \nonumber
\end{equation}
Then invertible dynamical map $\Lambda_t$ is CP-divisible if and only if
\begin{equation}\label{}
  \frac{d}{dt}\, || [{\rm id} \otimes \Lambda_t](X)||_\Box \leq 0 .
\end{equation}
\end{proposition}

In general, for $k$-divisible maps, the following holds true.

\begin{proposition}[\cite{PRL-kop}] If a dynamical map $\Lambda_t$ is $k$-divisible, then
\bea
\frac{d}{dt}\, p_{\rm guess}^{(\ell)}(\{p_i,\Lambda_t \circ \mathcal{E}_i\}) \leq 0 ,
\eea
for any $\ell \leq k$.
\end{proposition}

The result can equivalently stated in terms of the distance measure as follows.

\begin{proposition}[\cite{PRL-kop}] If a dynamical map $\Lambda_t$ is invertible, then it is $k$-divisible if and only if
\bea
 \frac{d}{dt} D^p_k(\mathcal{E}_1,\mathcal{E}_2)  \leq 0 \ ,
\eea
for all pairs of quantum channels $\mathcal{E}_{1}$ and $\mathcal{E}_{2}$, and arbitrary probability vector $(p_1,p_2)$.
\end{proposition}

To discriminate between quantum channels \cite{Gil}  we may use the operational fidelity defined in \cite{Belavkin} (see also \cite{Werner}). Given two channels $\mathcal{E}_1, \mathcal{E}_2 : \mathcal{B}(\mathcal{H}_A) \to \mathcal{B}(\mathcal{H}_B)$ one defines

\begin{equation}\label{}
  F(\mathcal{E}_1,\mathcal{E}_2) = \inf_{||\psi||=1} F([{\rm id}_A \otimes \mathcal{E}_1](|\psi\>\<\psi|), [{\rm id}_A \otimes \mathcal{E}_2](|\psi\>\<\psi|)) ,
\end{equation}
where the minimum is over all normalized vectors $\psi \in \mathcal{H}_A \otimes \mathcal{H}_A$. One proves the analog of (\ref{<<})

\begin{equation}\label{>>}
  1 - F(\mathcal{E}_1,\mathcal{E}_2) \leq \frac 12 ||\mathcal{E}_1 - \mathcal{E}_2||_\diamond  \leq \sqrt{  1 - F^2(\mathcal{E}_1,\mathcal{E}_2) } ,
\end{equation}
for any pair of states $\mathcal{E}_1$ and $\mathcal{E}_2$.

\begin{proposition} If $\Lambda_t$ is CP-divisible, then
\begin{equation}\label{}
  \frac{d}{dt} \, F(\Lambda_t \circ \mathcal{E}_1,\Lambda_t \circ \mathcal{E}_2) \geq 0 ,
\end{equation}
for any pair of states $\mathcal{E}_1$ and $\mathcal{E}_2$.
\end{proposition}

\section{Conclusions}  \label{CON}

In this paper, we have analyzed well known  concepts of quantum Markovianity based on CP- and P-divisibility of the corresponding dynamical maps.

The memoryless dynamics in classical information processing is characterized by the Markov process, which is governed by the classical conditional probability.
Its quantum analogue is, however, not uniquely defined, due to the lack of a unique definition of the quantum conditional state.
Various concepts and methods have been developed recently to deal with the somewhat elusive notion of quantum (non-)Markovian evolution~\cite{NM1,NM2,NM3,WISE}.
In particular, in~\cite{WISE} various approaches have been compared.

Here, we have first focused on a purely mathematical property of dynamical maps --- i.e., divisibility --- regarded as
a suitable notion of quantum Markovianity, and we have then related it to some physical properties, like distinguishability of states and channels.
This approach  has a direct connection with quantum information theory, where one analyzes various properties of quantum channels,
and of quantum operations in general.

The original proposal by Breuer~\emph{et al.}~\cite{BLP} introduces a concept of information flow and links it to quantum (non-)Markovianity.
This concept is strictly related to distinguishability of quantum states and its temporal evolution.
In this paper, we use instead other information based quantities, namely, various notions of relative entropy (or, in general, the so-called divergences~\cite{Hiai}).
If such a quantity satisfies the data processing inequality, then it may be used to witness non-Markovianity of the quantum evolution or, equivalently,
the lack of CP-divisibility of the corresponding dynamical map. Clearly, the fact that many relative entropies considered in the paper satisfy the data processing inequality
is well known. What seems to be new is the observation that this property may be used for witnessing quantum non-Markovianity.
Moreover, we have been able to generalize some of these witnesses in order to characterize a more refined property, that is, the $k$-divisibility of the dynamical map
(which may be considered as a weaker version of Markovianity~\cite{PRL-k}).

Interestingly, due to recent results in~\cite{Reeb}, there is a nice relation between the information flow of Breuer~\emph{\it et al.}
and the information flow based on the relative entropy: for any P-divisible map one has the monotonicity of $||\Lambda_t(\rho_1-\rho_2)||_1$.
However, the monotonicity of $D(\Lambda_t(\rho_1)||\Lambda_t(\rho_2))$ was only known to hold for maps which are at least 2-divisible.
Only recently it has been shown that P-divisibility is also sufficient~\cite{Reeb}. Notice that, in the case of the information flow,
one can easily prove the following: if the map $\Lambda_t$ is invertible, then the monotonicity of $||\Lambda_t(X)||_1$, for any Hermitian $X$,
implies that $\Lambda_t$ is P-divisible. Whether such a property holds for relative entropy based quantities, as well, is not known.

Very recently, the problem of non-invertible maps has been treated in~\cite{Erling}. Here, there is an essential difference between CP and P-divisibility.
This is due to the fact that, contrary to the case of CP maps, very little is known about positive extensions of positive maps
that are defined on linear subspaces of $\BH$ only. In the case of completely positive maps, one has the celebrated Arveson extension theorem,
and its generalization~\cite{Paulsen}. It is not clear whether invertibility plays any role in the case of entropy based quantities.

Finally, it is also interesting that the family of sandwiched R\'enyi divergences gives rise to the concept of min- and max- conditional entropy,
with a clear operational meaning. This leads to an appropriate operational meaning of quantum Markovianity, as well.
Therefore, the purely mathematical concept of CP-divisibility acquires a clear operational meaning in terms of monotonicity
of \emph{quantum correlations} and \emph{quantum decoupling}.
This approach may be immediately generalized to the family of smooth entropies~\cite{TOM}, that turned out to be a very effective tool in quantum information theory.
In this way, one may devise tools to witness non-Markovianity in the one shot scenario.

\section*{Acknowledgments}

DC was partially supported by the National Science Center project 2015/17/B/ST2/02026.  JB was supported by National Research Foundation of Korea (NRF-2017R1E1A1A03069961), ITRC program(IITP-2018-2018-0-01402), and the KIST Institutional Program (2E26680-18-P025).

\end{document}